# ANALYSIS AND IMPROVEMENT OF PAIRING-FREE CERTIFICATE-LESS TWO-PARTY AUTHENTICATED KEY AGREEMENT PROTOCOL FOR GRID COMPUTING

Amr Farouk, Mohamed M. Fouad and Ahmed A. Abdelhafez

Department of Computer Engineering, Military Technical College, Cairo, Egypt

***ABSTRACT***

*The predominant grid authentication mechanisms use public key infrastructure (PKI). Nonetheless, certificate-less public key cryptography (CL-PKC) has several advantages that seem to well align with the demands of grid computing. Security and efficiency are the main objectives of grid authentication protocols. Unfortunately, certificate-less authenticated key agreement protocols rely on the bilinear pairing, that is extremely computational expensive. In this paper, we analyze the recently secure certificate-less key agreement protocols without pairing. We then propose a novel grid pairing-free certificate-less two-party authenticated key agreement (GPC-AKA) protocol, providing a more lightweight key management approach for grid users. We also show, a GPC-AKA security protocol proof using formal automated security analysis Sycther tool.*

## 1. INTRODUCTION

Authentication in grid computing is considered as the first defense point of security aspects. Grid authentication (GA) mechanism has to be efficient and secure. Moreover, GA mechanism should fulfil the requirements of large scale distributed and heterogeneous grid virtual organizations (VO), that usually spans multiple trust domains. Due to the dynamic nature of grid VOs and frequent user requests, the identity of grid principals is verified by these authentication protocols using cryptographically secure mechanisms, that have the following challenges: i) mutual authentication for multi-domain, scalability and delegation, ii) single sign-on, identity federation and credentials confidentiality, and iii) efficiency, lightweight and flexibility [1].

The most prevalent grid security standard, grid security infrastructure (GSI) uses the SSL authentication protocol (SAP) to achieve mutual entity authentication between user proxy (UP) and resource proxy (RP) [2]. Hence, SAP rely on offline certificate-based public key authentication infrastructure (e.g., X.509), that bring about problems to certificates management hindering grid scalability, such as poor inter-operability in hierarchical PKIs, certificate revocation, and poor usability. Therefore, the certificate-free authentication has emerged based on identity-based cryptography (IBC) using pairings. In IBC, an entity's public key is directly derived from its identity information (e.g., name, e-mail address, telephone number, and IP address). Generally, identity-based authentication overcomes such aforementioned problems with its certificate-free feature, however, the private key generator (PKG) used for key distribution has a key escrow shortcoming. Recently, Wang et. al. [2] present the first grid certificate-less authentication based on certificate-less public key cryptography (CL-PKC), that is a kind of cryptography between certificate-based and identity-based PKC. Al-Riyami and Paterson [3] introduce the CL-PKC to solve the key escrow problem imposed in IBC. CL-PKC not only preserves all the advantages of IBC, but also avoids the use of certificates. As well, CL-PKC uses a trusted authority called key generation center (KGC). The later only generates a partial private key given the users' identity. The user's private key is generated by both a partial private key and a secret value chosen by the user. The user's public key does not need to be certified, so it is implicitly certified by the partial private key issued by the KGC.

All the grid authentication mechanisms mentioned above, either have a security issues or are not efficient to be practical implemented in real environments.

This paper introduces the first pairing-free CL-AKA protocol used in grid computing environment. We propose a novel secure and efficient pairing-free CL-AKA for grid authentication (GPC-AKA). Moreover, we provide a security proof for GPC-AKA using an automated security protocols verification tool.

The rest of this paper is organized as follows. A literature survey is presented on pairing-free CL-AKA protocols in Section 2. CL-PKC fundamental concepts are described in Section 3. Section 4 shows the AKA adversary models. Recently pairing-free CL-AKA protocols security analysis are introduced in Section 5. Our proposed grid pairing-free CL-AKA is presented in Section 6. Finally, we draw our conclusions in Section 7.

## 2. Literature Review

The key agreement (KA) is one of the fundamental cryptographic primitives in public key cryptography. KA allows a shared secret, called a session key intended for cryptographic use, to be available for two or more parties. For instance, if entity *A* is assured that no other entity besides entity *B* can possibly expose the secret key value, a key agreement protocol is said to provide implicit key authentication of entity *B* to *A*. A key agreement protocol, that provides mutual implicit key authentication, is called an authenticated key agreement (AKA) protocol [4]. There are three types of two-party AKA protocols, considering the message exchange during the protocol: i) *one-round or two pass*: both entities require to transmit information to each other, ii) *one-way*: only one entity is required to transmit information to the other, and iii) *non-interactive*: no information needs to be transmitted between two entities.

For fully secure and efficient grid entities authentication, it is required to build AKA protocol with high security. However a minimal number of communication passes and low computation cost. Certificate-less authenticated key agreement (CL-AKA) protocols is based on bilinear pairings which relative computation cost is approximately twenty times higher than that of a scalar multiplication over elliptic curve group [5]. The pairing is then considered as an expensive cryptography primitive. Therefore, several CL-AKA protocols, without pairing, have been proposed to improve efficiency as shown in Table 1.

From the preceding review, we focus on the more recent secure pairing-free CL-AKA protocols. Although the protocol of Yang et. al. [7] is provably secure, it's a highly computationally cost as mentioned later in Section 6.3. The proposed Debiao et. al. [10] and Nashwa et. al. [5] secure protocols, will be analyzed in Section 5.

Table 1. CL-AKA without paring protocols.

| Protocol | Cryptography Problem[1] | Security Model[2] | Security Weakness |
|---|---|---|---|
| Geng et. al. [4] | GDH | Random oracle | Type 1 adversary |
| Hou et. al. [6] | GDH | mBR | Type 1 adversary |
| Yang et. al. [7] | GDH | eCK | Provably secure |
| He et. al. [8] | GDH | mBR | Type 1 adversary |
| He and chen [9] | GDH | Random oracle | Type 1 adversary |
| Debiao et. al. [10] | GDH | eCK | Provably secure |
| Nashwa et. al. [5] | BDH | Swanson | Secure |

[1] Will be discussed in Section 3.3

[2] Will be discussed in Section 4

# 3. PRELIMINARIES OF CL-PKC

This section shows the cryptography notations and fundamentals required to grasp CL-PKC, as Section 3.1 introduces elliptic curve cryptography. Section 3.2 presents the bilinear pairing computation process, then Section 3.3 shows the cryptography computational-based problems. Riyami and Paterson CL-AKA scheme background is reviewed in Section 3.4.

## 3.1. Elliptic Curve Cryptography

Elliptic curve cryptosystems have the potential to provide relatively small block size, high-security public key schemes that can be efficiently implemented [11]. Let $E/F_p$ denote an elliptic

curve, $E$, over a prime finite field $F_p$, in (1)

$$y^2 = x^3 + ax + b, \qquad \forall a, b \in F_q, \tag{1}$$

and with the discriminant defined by (2)

$$\Delta = 4a^3 + 27b^2 \neq 0 \tag{2}$$

The points on the curve $E/F_p$ together with an extra point $O$, called the point at infinity, form a group $G = \{(x, y) : x, y \in F_p, E(x, y) = 0\} \cup \{O\}$. $G$ is a cyclic additive group in the point addition "+" defined as follows: Let $P,Q \in G$, $l$ be the line containing $P$ and $Q$ (tangent line to $E/F_p$ if $P = Q$), and $R$, the third point of intersection of $l$ with $E/F_p$. Let $l'$ be the line connecting $R$ and $O$. Then $P$ "+" $Q$ is the point, such that $l'$ intersects $E/F_p$ at $R$ and $O$. Scalar multiplication over $E/F_p$ can be computed as follows: $tP = P + P + \cdots + P$ ($t$ times).

## 3.2. Bilinear Pairing

Let $G_1$ and $G_2$ be additive and multiplicative cyclic groups of prime order $q$, respectively. Let $P$ denote a generator in $G_1$. A bilinear pairing is a map $e : G_1 \times G_1 \rightarrow G_2$ that satisfies the following properties [12]:

1. Bilinearity: $e(aP, bQ) = e(P,Q)^{ab}$ $\forall P,Q \in G_1$ and $a, b \in Z_q$.
2. Non-degeneracy: $P \neq 0 \Rightarrow e(P, P) \neq 1$.
3. Computability: $e$ is efficiently computable.

The above properties also imply $\forall P,Q,R \in G_1$: $e(P+Q,R) = e(P,R).e(Q,R)$, $e(P,Q+R) = e(P,Q).e(P,R)$. Typically, the map $e$ will be derived from either the Weil or Tate pairing on an elliptic curve over a finite field.

### 3.3. Computational-based Problems

The security of certificate-less cryptography protocols is based on some well-studied problems that are assumed to be hard to compute efficiently. For appropriately selected parameters, the following problems are computationally intractable [12]:

1. Computatinal Diffie Hellman (CDH): given $P$, $aP$, $bP \in G_1$, compute $abP \in G_1$.
2. Decisional Diffie Hellman (DDH): given $P$, $aP$, $bP$, $cP \in G_1$, decide whether $c = ab \pmod q$ or not.
3. Gap Diffie Hellman (GDH): given $P$, $aP$, $bP \in G_1$, and DDH oracle, compute $abP$.
4. Bilinear Diffie-Hellman (BDH): given $P$, $aP$, $bP$, $cP \in G_1$, compute $e(P, P)^{abc} \in G_2$, where $e$ a bilinear pairing on $(G_1,G_2)$.

### 3.4. Riyami and Paterson CL-AKA Scheme Background

Al-Riyami and Paterson [3] introduce the concept of certificate-less public key cryptography (CL-PKC). The two-party CL-AKA scheme consists of two phases. First is the *setup phase*, runs between KGC and entities. It consists of five probabilistic polynomial time (PPT) algorithms: Setup, SetSecretValue, SetPartialPrivateKey, SetPrivateKey, SetPublicKey. Second is the *key agreement* phase. It runs between two entities, by PPT interactive algorithm: SessionKeyAgreement.

In fact, all the successive CL-AKA protocols that improve the security and efficiency, are based on Al-Riyami and Paterson protocol.

## 4. CL-AKA Protocols Adversary Models

This section discuss the most robust adversary models for CL-AKA protocols. Table 2 shows the general security attributes that apply to key agreement protocols.

There are two types of adversaries, namely, type I adversary ($A_I$ ) and type II adversary ($A_{II}$ ) with different capabilities in CL-PKC. $A_I$ does not have access to the KGC master-key, but has the ability to replace the public key of any entity with a value of its choice (i.e., acts as a dishonest user). Whereas, $A_{II}$ has an access to the KGC master-key, but cannot replace participants' public keys (i.e., acts as a malicious KGC from the beginning of the system setup or an honest-but-curious KGC, that is malicious after it has generated a master public/secret key pair honestly) [4].

The foremost adversarial model used for AKA protocols security proof is the extended Canetti and Krawczyk (eCK) model [13]. This model captures unknown key-share (UKS) and Key-compromise impersonation (KCI) attacks, however, it does not capture perfect forward secrecy (PFS). Swanson [14] propose the first formal security model for the CL-AKA protocol, however, it is a weak security model as $A_I$ is not allowed to replace the public key associated with the challenge identity. Lippold et. al. [15] transform original eCK model from the traditional PKI-based setting to the CL-PKC setting. Its strength comes from the ability of an user to use the new public/private key pair in the rest of the game after the adversary replaces the public key of the user. Nevertheless, this model is suitable for key agreement protocols based on pairing.

Table 2. Key agreement protocols security attributes.

| **Security Attributes** | **Security Credentials** | |
|---|---|---|
| | **Compromised** | **Should not be affected (By Adversary)** |
| Known-key secrecy | Other rounds session keys | Session key generated |
| Forward secrecy | One or more entities's long term private keys | Session keys generated |
| Perfect forward secrecy (PFS) | All entities's long term private keys | Session keys generated |
| KGC forward secrecy | KGC's master key Session keys generated KGC's master key | Session keys generated |
| Key-compromise impersonation resilience (KCI) | Entity A's long term private key | Session key with A by acting as another entity B |
| Unknown key-share resilience (UKS) | Entity A should not share a key with entity C | When in fact A thinks that it is sharing the key with entity B |
| No key control | Session key should be determined jointly by both entities | None of the entities can control the key alone to be the preselected values |
| Known session-specific temporary information security | Randomized input used in a protocol run | Session keys generated |

The aforementioned CL-AKA adversary models are defined via a simulation game between a challenger $C$ and an adversary $A_I$ or $A_{II}$ . The adversary inquires a polynomial number of queries, such as Create User, Send, Reveal MasterKey, RevealPartialKey, Reveal Secret Value, Reveal Ephemeral Key, ReplacePublicKey, RevealSessionKey, and Test. At the end of the game, the adversary $A_I$ or $A_{II}$ outputs a bit, $b'$, as its guess for random coin, $b$, flipped by $C$. The $A_i$'s advantage of winning the game can be defined as (3),

$$Adv(A_i) = | Pr(b = b') - 1/2 |, \quad \forall i \in \{I, II\} \tag{3}$$

## 5. PAIRING-FREE CL-AKA PROTOCOLS SECURITY ANALYSIS

A formal security proof for most CL-AKA protocol haven't been considered. Some other CL-AKA protocols are also proposed with heuristic security analysis [9]. In Section 5.1 and Section 5.2, we explore more recent robust pairing-free CL-AKA protocols and address its security features to improve efficiency and security of pairing-free CL-AKA protocol for grid computing environment.

### 5.1. Nashwa et. al. Scheme

Nashwa et. al. [5], introduce a fully secure and efficient CL-AKA without pairing into two phases, as illustrated in Figure 1 and Figure 2, respectively.

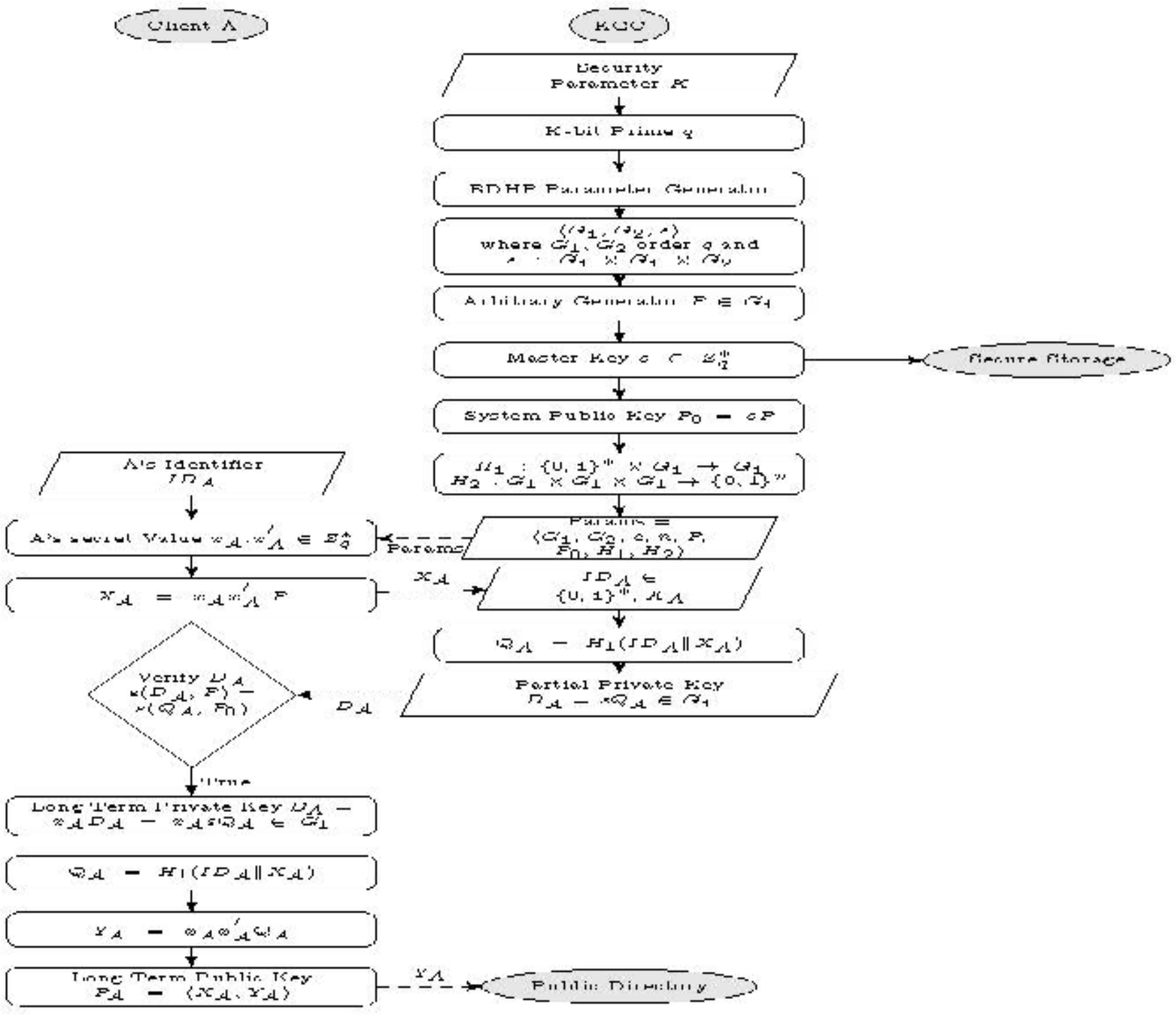

Figure 1. Nashwa et. al. [5] key generation setup scheme (Phase 1)

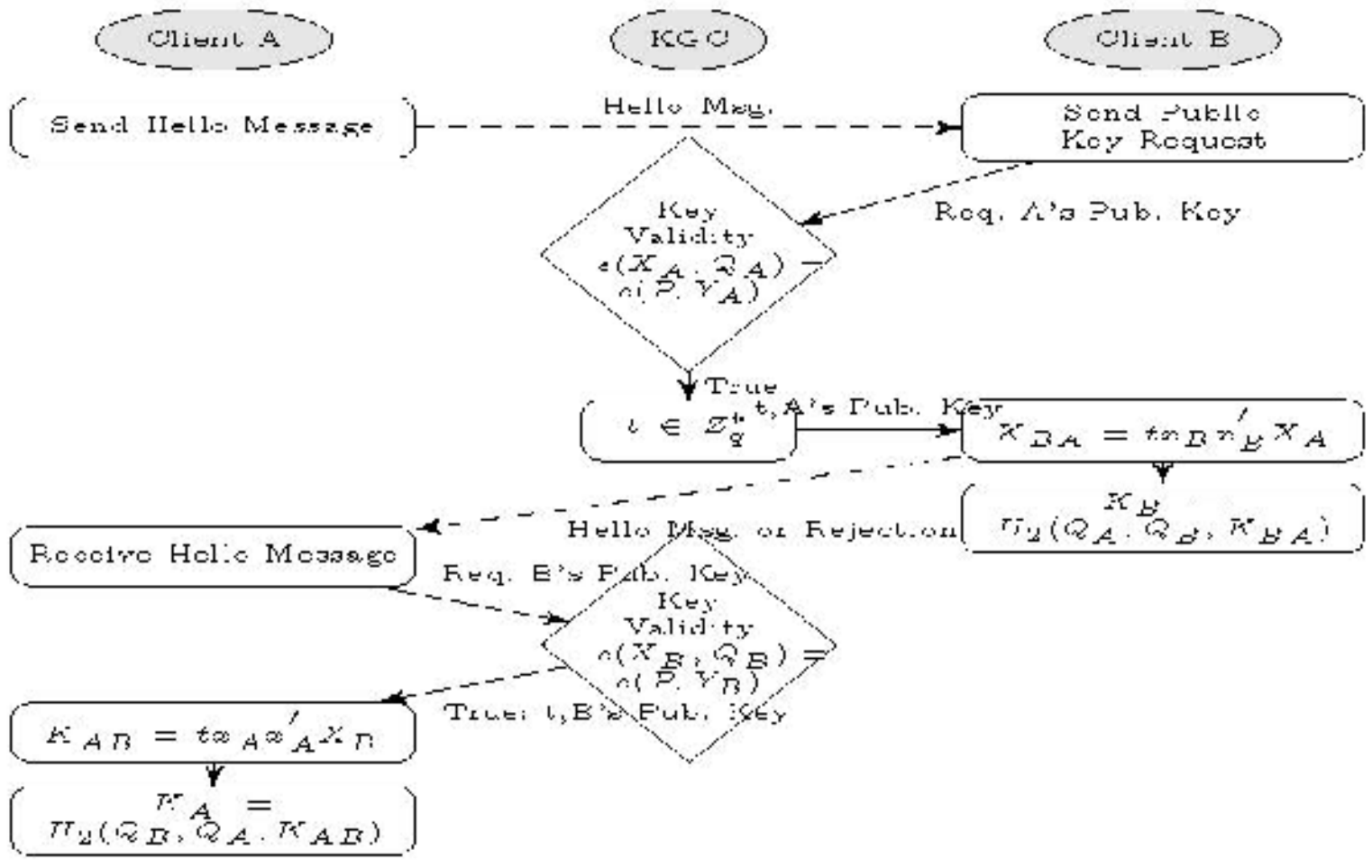

Figure 2. Nashwa et. al. [5] key agreement scheme (Phase 2).

This protocol isn't pure pairing-free CL-AKA because they move the pairing from entities to KGC. In the protocol setup phase, shown in Figure 1, there are two pre-computed bilinear pairings run at the user side to verify the partial private key, $D_A$, computed at KGC. In the key agreement phase, shown in Figure 2, KGC performs one pairing for each entity to check whether the entity requested's public key is valid within the domain or not (i.e., verify that same KGC master key is used).

This protocol is non-interactive key agreement protocol (i.e., KGC is involved in key agreement phase) yielding some drawbacks, such as: i) entities do not authenticate KGC to prevent $A_{II}$ , ii) the KGC is a single point of failure and vulnerable to DOS attacks or incomplete public key requests, and iii) the scalability problem rises in a large scale distributed environments, such as grid computing.

Nashwa et. al. [5] claim that their proposed protocol is fully secure against $A_I$ and $A_{II}$ , and provably secure against Swanson security attributes model. However, we perform a security analysis in six points as follows:

1. **Known-key secrecy:** This comes from the fact that, each run of the protocol between two entities $A$ and $B$, a random, $t$, is selected from KGC and no other entities's ephemeral values (i.e., $t$ is the only fresh value) involved in session key agreement, so it is vulnerable to $A_{II}$ .
2. **Forward secrecy:** Outside adversary has compromised secret value of one entity or more $x_{ID}$, $x'_{ID}$ can not reveal the current session key $K$ due to the ephemeral random value $t$, however, a malicious KGC can determine the previously established session key. So, this protocol is not secure against $A_{II}$ and does not provide forward secrecy. Moreover, it can not achieve the perfect forward secrecy.
3. **KGC forward secrecy:** In this protocol, if the master key, $s$, of KGC is corrupted, the security of session keys previously established will not be compromised by any entity due to the ephemeral random value, $t$.
4. **Key-compromise impersonation:** It is possible for adversary, $E$, to impersonate any other entities like $B$ to $A$. E knows $S_A$ and get $t$, $A$'s public key, so $E$ can compute the shared secret key. Thus, it is vulnerable to $A_I$ (e.g., KCI).
5. **Unknown key-share resilience:** Entities like $A$ uses $Q_A$ in computing partial public key $Y_A$ for public key authenticity. Only KGC who checks the entities's public key authenticities, so this protocol is vulnerable to $A_I$ (e.g., public key replacement).
6. **No key control:** Shared session key should be computed with parameters from all entities, and no session key control. However the KGC is the only one originate the random value, $t$. So, in $A_{II}$ , malicious KGC can predetermine session key $K$, (e.g., key escrow).

According to the aforementioned security analysis of Nashwa et. al. protocol [5], it can be shown that, it is not secure against $A_I$ and $A_{II}$ . Therefore, it is inappropriate for grid computing authentication.

## 5.2. Debiao et. al. Scheme

Debiao et. al. [10], introduce a provably secure CL-AKA without pairing, as shown in Figure 3 and Figure 4 illustrating the key generation setup, and the key agreement schemes, respectively.

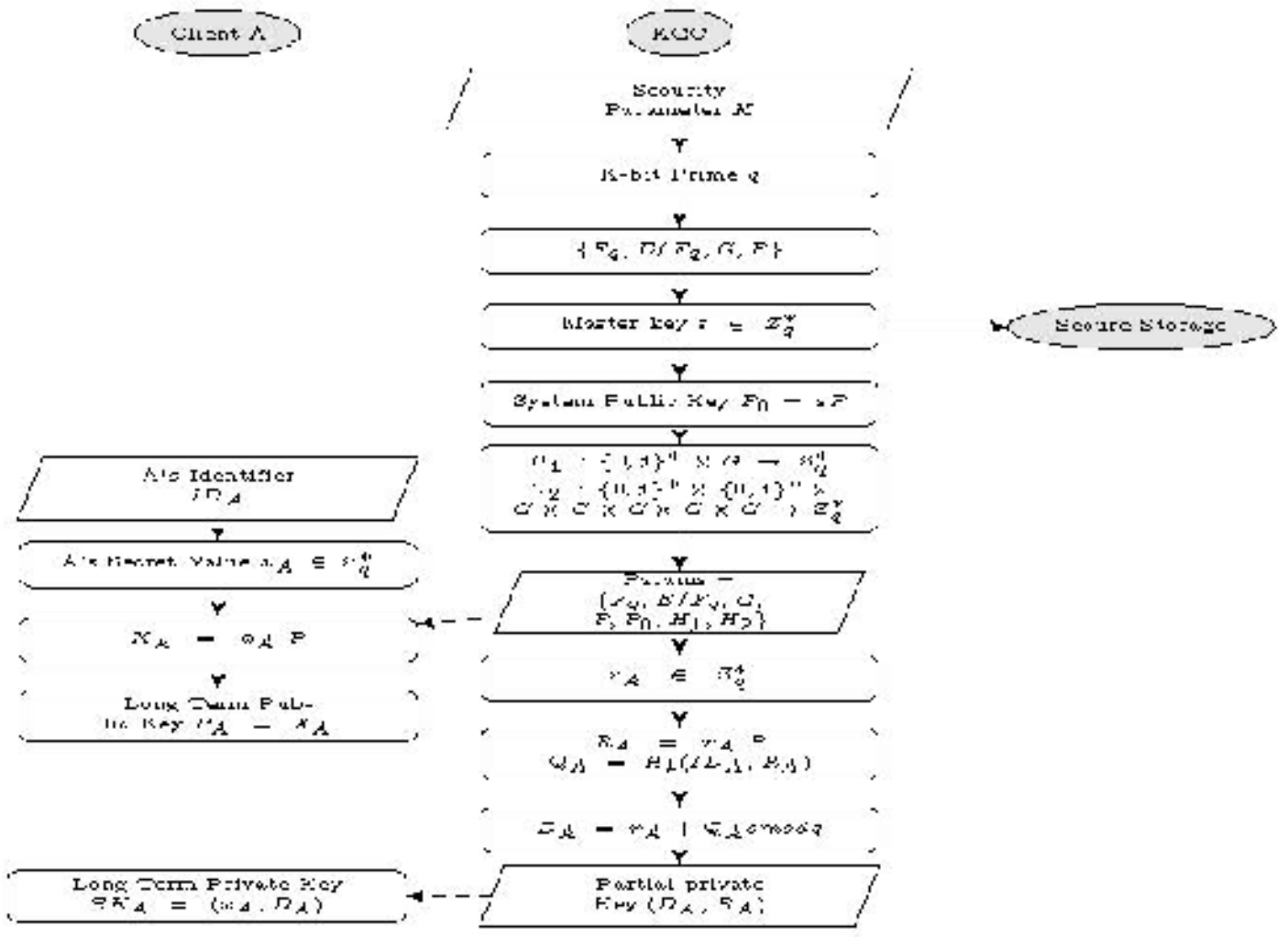

Figure 3. Debiao et. al. [10] key generation setup scheme (Phase 1).

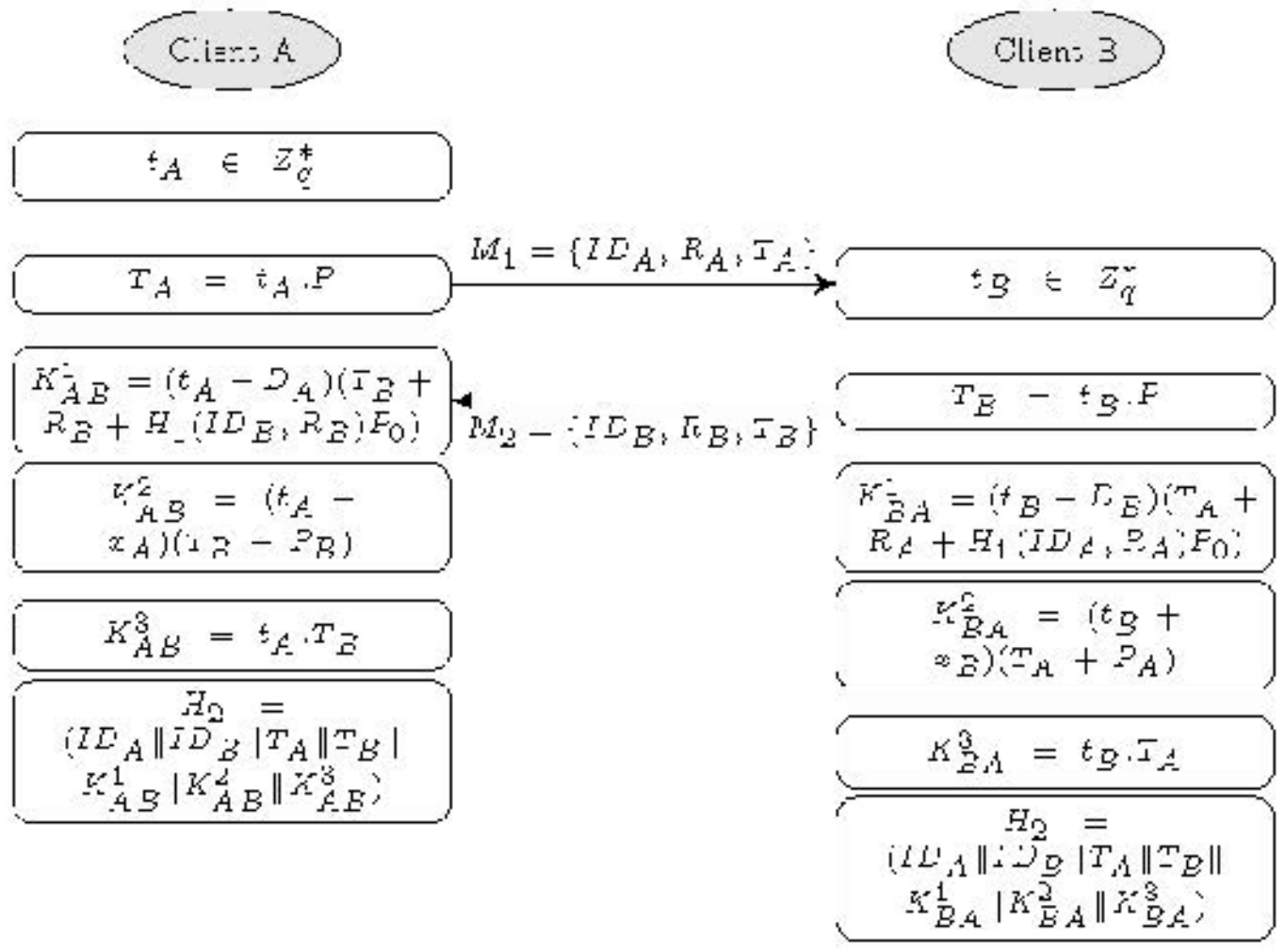

Figure 4. Debiao et. al. [10] key agreement scheme (Phase 2).

Debiao et. al. protocol [10] is provably secure against eCK adversary model. Moreover, we perform an analysis in six security attributes as follows:

1. **Known-key secrecy:** Each session key is unique due to the freshness of $A$ and $B$ entities's ephemeral values $t_A$, $t_B$ respectively. Thus, compromising a session key will not affect past or future sessions.
2. **Forward secrecy:** Even if the long-term private key(s) is compromised, $A_I$ does not reveal previously established session keys without the knowledge of $t_A$, $t_B$, that is exactly a CDH problem.
3. **KGC forward secrecy:** In CL-PKC based schemes, if the KGC's master secret key is compromised, the previously established session keys will not be exposed.

4. **Key-compromise impersonation:** If $A_I$ compromise the $A$ entity's long-term private key, he would be unable to compute the session key, as $x_A$ is unknown.
5. **Unknown key-share resilience:** Each entity implicit authenticate who it shares the secret key with, as $R_A$,$R_B$ are used for computing the session key.
6. **No key control:** As most of the existing two-party AKA protocols, the responder entity gains an unfair advantage over the initiator entity.

We can infer from the above security analysis that Debiao et. al. protocol [10], is secure against $A_I$ and $A_{II}$. So, it is appropriate for grid computing environment. In the next section, we propose a novel grid pairing-free CL-AKA protocol to improve the performance of Debiao et. al. protocol [10], as well as we present a security proof for the proposed protocol.

## 6. PROPOSED GRID PAIRING-FREE CL-AKA

In Section 6.1, we present the first secure and efficient pairing-free certificate-less two-party authenticated key agreement protocol for grid computing (GPC-AKA). Furthermore, a security proof using Scyther tool is provided in Section 6.2. Our proposed scheme is more efficient than those provable secure schemes as explained in Section 6.3.

### 6.1. Proposed Protocol (GPC-AKA)

Our proposed secure and efficient GPC-AKA protocol for grid computing has the following properties: i) pairing-free, less computational cost so to be more efficient. ii) two-party, due to grid computing environment scalability and dynamic features. iii) one-round, less network overload so more efficient. iv) the proxies (i.e., UP, RP) can help to meet frequent mutual authentication requests between users and resources, so support grid single sign on (SSO). v) unique node's registered distinguished name (DN) from root to node, to provide cross-trust domain in which each domain comprises one KGC. Before authentication, trust relationship has built between KGCs to shared system parameters with each other.

Based on Debiao et. al. protocol we make the following modification to achieve more efficient and secure protocol used for grid computing environment: i) In phase 1, we use the hashing functions

$H_1 : \{0, 1\}^{*} \times G \times G \rightarrow Z^{*}_q$
$H_2 : \{0, 1\}^{*} \times \{0, 1\}^{*} \times G \times G \times G \rightarrow Z^{*}_q$

ii) In phase 2, We have embedding the entity long term public key $P_A$ in key derivation function $H_1$.

Our proposed Pairing-free certificate-less two party authenticated key agreement for grid (GPC-AKA) is depicted in Figure 5.

**GPC-AKA protocol consistency is proved:**

$$K_{AB} = (t_A + D_A + x_A)(T_B + P_B + R_B + H_1(ID_B,R_B, P_B)P_0)$$
$$= (t_A + D_A + x_A)((t_B.P) + (x_B.P) + (r_B.P) + (Q_B.sP))$$
$$= (t_A + D_A + x_A)(t_B + x_B + r_B + Q_B.s)P$$
$$= (t_A + D_A + x_A)(t_B + x_B + D_B)P = K_{BA}$$

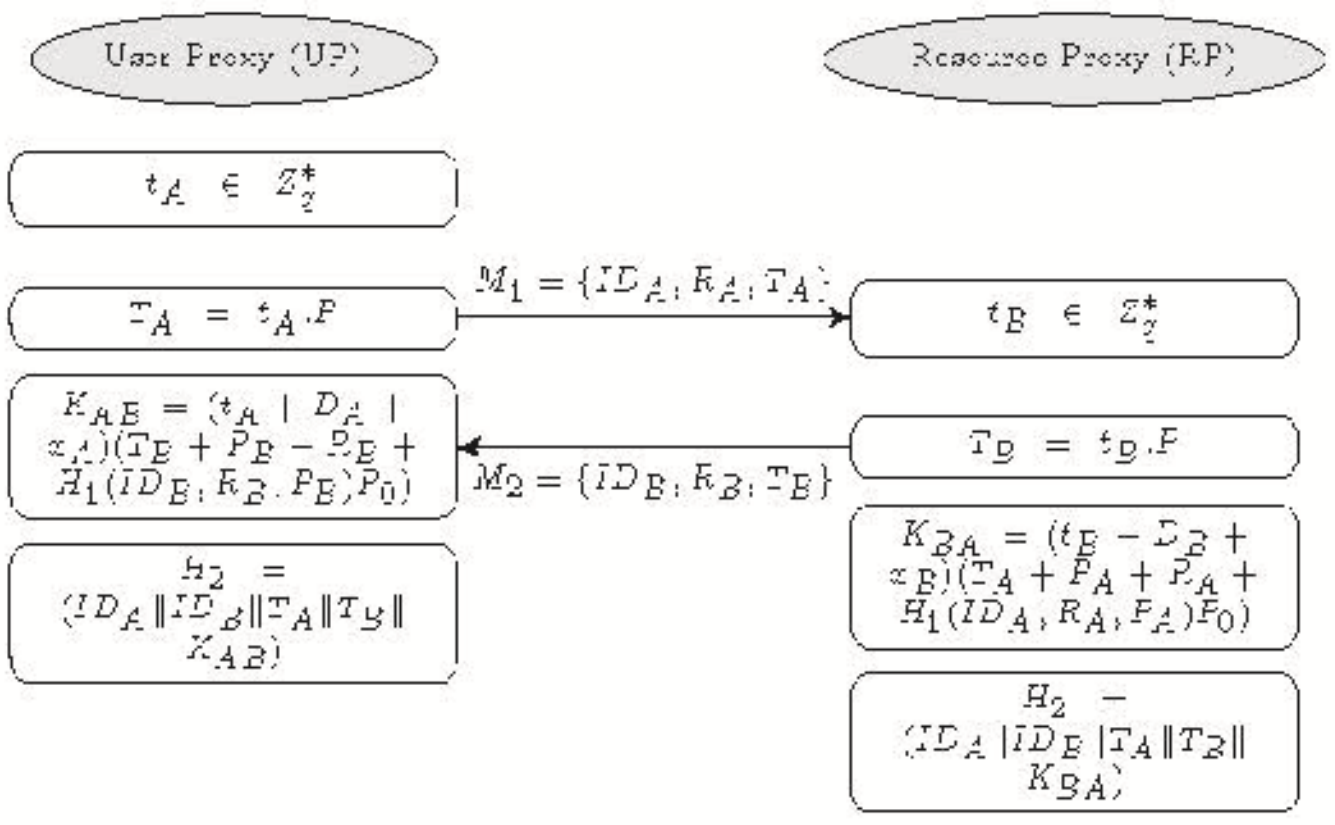

Figure 5. Proposed key agreement scheme GPC-AKA (Phase 2).

## 6.2. Security Proof

We provide a formal security analysis, using automated security protocol verification tool Scyther version compromise-0.8, on laptop 2 GHz Intel core 2 duo processor, with 6 GB RAM. Scyther tool present a framework for modeling adversaries in security protocol analysis, ranging from a Dolev-Yao style adversary to more powerful adversaries, supports notions such as weak perfect forward secrecy, key compromise impersonation, and adversaries capable of state-reveal queries [16].

Figure 6 shows the settings of the adversary model used in verifying our proposed GPC-AKA protocol.

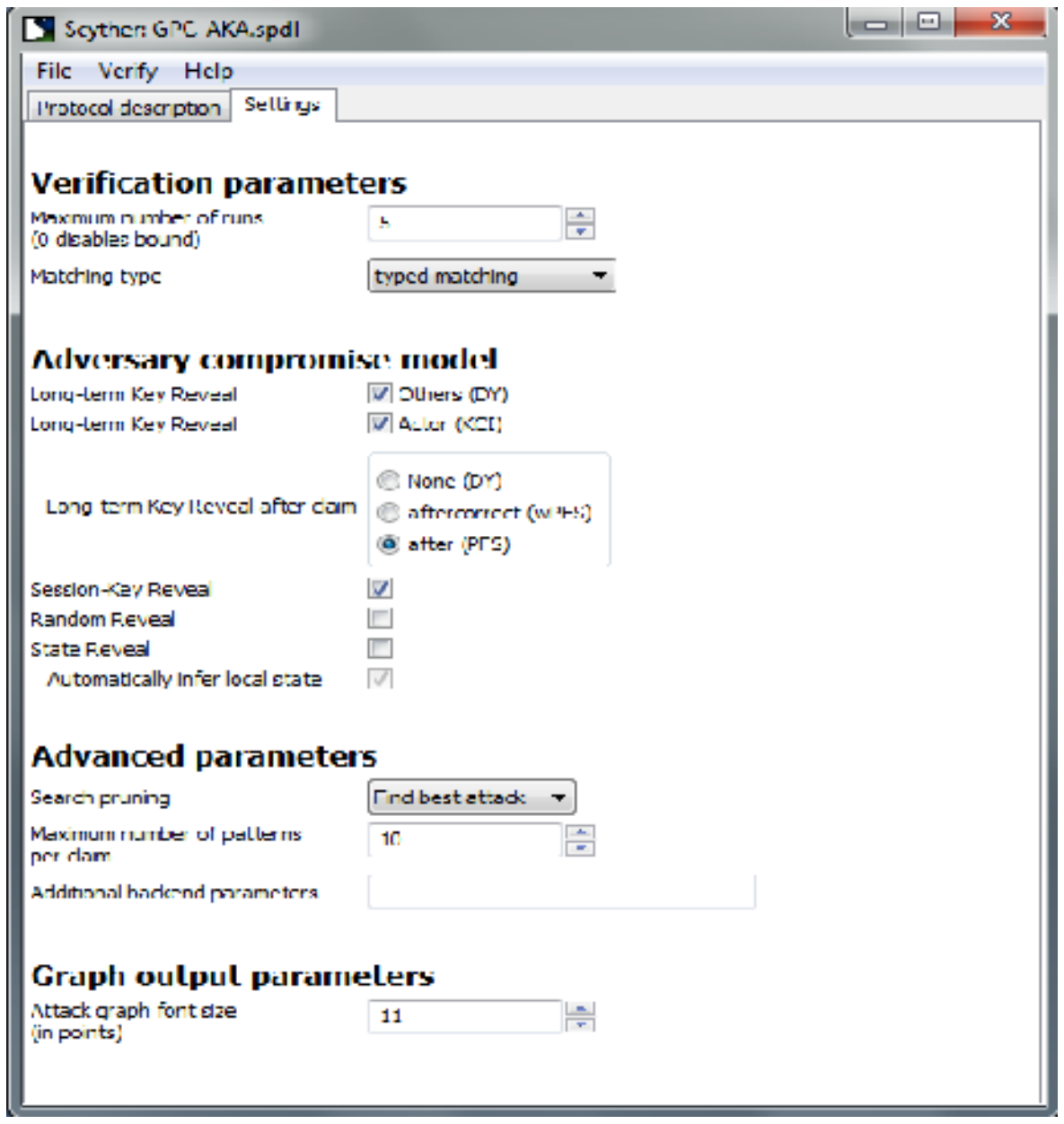

Figure 6. Scyther adversary model used for GPC-AKA verifying.

We model the GPC-AKA protocol in security protocol description language (SPDL) using Sycther tool as follows:

```
/*
* Proposed Pairing-free CL-AKA for Grid (GPC-AKA)
*/
// Hash functions
hashfunction KDF,H;
// Addition, multiplication, simply hashes
hashfunction mult,add;
// The protocol description
protocol GPC-AKA(KGC,UP,RP)
// UP = Initiator, RP = Responder
{
const rA,rB,P;
role KGC // Key Generation Center
{
send_1(KGC,UP,P); // Publish public params
send_2(KGC,RP,P);
}
role UP // User Proxy
{
fresh tA: Nonce; // Ephemeral Secret
var TB: Ticket;
var RB;
recv_1(KGC,UP,P);
send_3(UP,RP,mult(rA,P)); // Send RA
recv_4(RP,UP,RB);
send_5(UP,RP, mult(tA,P)); // Send TA
recv_6(RP,UP,TB);
// Secret Session Key
claim(UP,SKR,KDF(UP,RP,mult(tA,P),TB,mult
(add(tA,sk(UP,KGC),sk(UP)),add(TB,pk(RP),
RB,mult(H(RP,RB,pk(RP))),pk(KGC)))));
}
role RP // Resource Proxy
{
fresh tB: Nonce; // Ephemeral Secret
var TA: Ticket;
var RA;
recv_2(KGC,RP,P);
recv_3(UP,RP,RA);
send_4(RP,UP,mult(rB,P)); // Send RB
recv_5(UP,RP,TA);
send_6(RP,UP,mult(tB,P)); // Send TB
// Secret Session Key
claim(RP,SKR,KDF(UP,RP,TA,mult(tB,P),mult
(add(tB,sk(RP,KGC),sk(RP)),add(TA,pk(UP),
RA,mult(H(UP,RA,pk(UP))),pk(KGC)))));
}
}
```

Figure 7 shows the proposed GPC-AKA verification using Scyther tool.

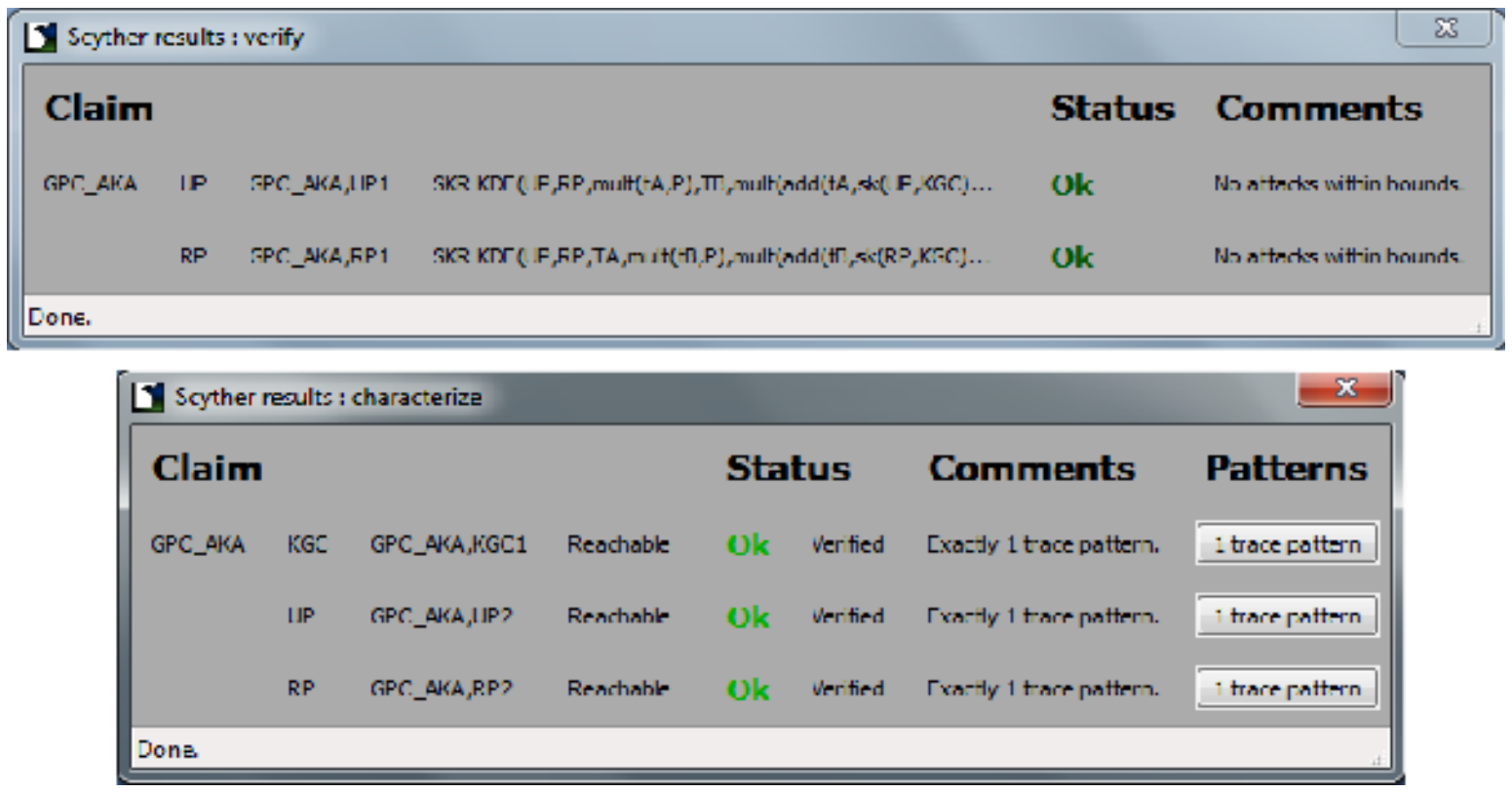

Figure 7. GPC-AKA scyther security protocol verification.

## 6.3. Performance Analysis

We compare the efficiency of our protocol GPC-AKA to Geng et. al. [4], Hou et. al. [6], Yang et. al. [7], He et. al. [8], He and chen [9], Debiao et. al. [10], and Nashwa et. al. [5], in terms of computational cost and communication overheads as shown in Table 3.

Table 3. Pairing-free CL-AKA protocols efficiency comparison.

| Protocol | Computational cost | | | | | Message exchange |
|---|---|---|---|---|---|---|
| | $T_{BP}$ [1] | $T_{inv}$ [2] | $T_{mul}$ [3] | $T_{add}$ [4] | $T_h$ [5] | |
| Geng et. al. [4] | | | 7 | | 2 | 2 |
| Hou et. al. [6] | | | 6 | | 2 | 2 |
| Yang et. al. [7] | | | 9 | | 2 | 2 |
| He et. al. [8] | | 1 | 5 | 3 | 2 | 2 |
| He and chen [9] | | | 5 | 4 | 2 | 2 |
| Debiao et. al. [10] | | | 5 | 3 | 2 | 2 |
| Nashwa et. al. [5] | 1(KGC) | | 2 | | 1 | 1+2(KGC) |
| Proposed GPC-AKA | | | 3 | 5 | 2 | 2 |

[1] Bilinear pairing
[2] Multiplicative inverse
[3] Elliptic curve point multiplication
[4] Elliptic curve point addition
[5] Hashing

In this comparison we consider the computations for single entity in key agreement scheme, excluding the pre-computed ones. It can be shown that, our proposed protocol re- quires less computation cost, i.e., only 3 multiplication operation, as compared to other protocols that require 7, 6, 9, 5, 5, 5, and 2, operations respectively. As well, it can be noticed that the proposed protocol keeps the same communication overhead as compared to the others (i.e., only 2 message exchange). Hence the GPC-AKA protocol is more efficient and appropriate for a practical grid computing environment.

## 7. Conclusions

The certificate-less authenticated key agreement (CL-AKA) approach brought a significant impacts on grid computing authentication. Whereas CL-AKA implementations rely on bilinear pairing which is computationally expensive. Encouragingly, pairing-free CL-AKA can overcame this performance drawback, particularly in scalable and dynamic grid computing environment. In this paper, we present a security analysis for the recently secure CL-AKA without pairing, Nashwa et. al. and Debiao et. al. protocols. Unfortunately, Nashwa et. al. protocol is not secure against type I and type II adversaries. Based on Debiao et. al. protocol, a novel secure and efficient pairing-free two-party certificate-less authenticated key agreement protocol for grid computing (GPC-AKA) is proposed, which improve the performance. Our protocol is provably secure using a formal security protocol verification Sycther tool. Eventually, the performance analysis comparison indicates that our proposed protocol GPC-AKA is more efficient than previous pairing-free CL-AKA protocols.

## REFERENCES

[1] Amr Farouk, Ahmed A. Abdelhafez, and Mohamed M. Fouad. Authentication mechanisms in grid computing environment: Comparative study. In IEEE International Conference on Engineering and Technology, pages 1–6, Oct. 2012.

[2] Wang Shengbao, Cao Zhenfu, and Bao Haiyong. Efficient certificateless authentication and key agreement (CL-AK) for grid computing. In International Journal of Network Security, volume 7, pages 342–347, Nov. 2008.

[3] Al-Riyami Sattam and Paterson Kenneth. Certificateless public key cryptography. In Springer Journal on Advances in Cryptology-Asiacrypt, pages 452–473, 2003.

[4] Geng Manman and Zhang Futai. Provably secure certificateless two-party authenticated key agreement protocol without pairing. In IEEE International Conference on Computational Intelligence and Security, volume 2, pages 208–212, 2009.

[5] Mohamed Nashwa, Hashim Mohsin, Bashier Eihab, and Hassouna Mohamed. Fully-secure and efficient pairing-free certificateless authenticated key agreement protocol. In IEEE World Congress on Internet Security, pages 167–172, 2012.

[6] Hou Mengbo and Xu Qiuliang. A two-party certificateless authenticated key agreement protocol without pairing. In the 2nd IEEE International Conference on Computer Science and Information Technology, pages 412–416, 2009.

[7] Yang Guomin and Tan Chik-How. Strongly secure certificateless key exchange without pairing. In the 6th ACM Symposium on Information, Computer and Communication Security, pages 71–79, 2011.

[8] He Debiao, Chen Jianhua, and Hu Jin. A pairing-free certificateless authenticated key agreement protocol. International Journal of Communication Systems, pages 221–230, 2012.

[9] He Debiao, Chen Yitao, Chen Jianhua, Zhang Rui, and Han Weiwei. A new two-round certificateless authenticated key agreement protocol without bilinear pairings. Mathematical and Computer Modelling, 54(11):3143–3152, Aug. 2011.

[10] He Debiao, Padhye Sahadeo, and Chen Jianhua. An efficient certificateless two-party authenticated key agreement protocol. Computers & Mathematics with Applications, 64(6):1914–1926, Sep. 2012.

[11] J. Menezes Alfred, Okamoto Tatsuaki, and A. Vanstone Scott. Reducing elliptic curve logarithms to logarithms in a finite field. IEEE Transactions on Information Theory, 39(5):1639–1646, Sep. 1993.

[12] Saxena Amitabh and Soh Ben. Applications of pairings in grid security. In IEEE First International Conference on Communication System Software and Middleware, pages 1–4, 2006.

[13] Benedikt Schmidt. Formal Analysis of Key Exchange Protocols And Physical Protocols. Ph.d. thesis, University of Karlsruhe, Karlsruhe, Wurttemberg, Germany, Nov. 2012.

[14] Colleen Marie Swanson. Security in key agreement: Two-party certificateless schemes. Master thesis, Department of Mathematics, University of Waterloo, Waterloo, Ontario, Canada, 2009.

[15] Lippold Georg, Boyd Colin, and Nieto Juan. Strongly secure certificateless key agreement. In Springer Journal on Pairing-Based Cryptography, pages 206–230, 2009.

[16] Basin David and Cremers Cas. Modeling and analyzing security in the presence of compromising adversaries. In Springer Journal on Computer Security, pages 340–356, 2010.

## Authors

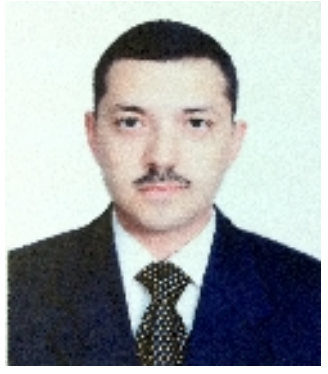

**Amr Farouk** received the Bachelor engineering from the Military Technical College (MTC), Cairo, Egypt, in 1997, and the Masters' engineering degrees from Engineering faculty, Mansoura university, Mansoura, Egypt in 2009. He is currently a PhD arguing from Computer engineering, MTC, Cairo, Egypt. His research interests include network security, authentication protocols, certificate-less authenticated key agreement.

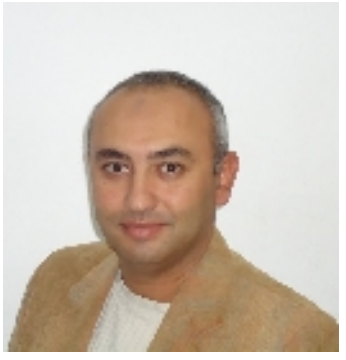

**M. M. Fouad** received the Bachelor engineering (honors, with great distinction) and Masters' engineering degrees from the Military Technical College (MTC), Cairo, Egypt, in 1996 and 2001, respectively. As well, he received the Ph.D. degree in Electrical and Computer engineering from Carleton University, Ottawa, Ontario, Canada, in 2010. He is currently a faculty member with the Department of Computer Engineering, MTC. His research interests are in online handwritten recognition, image registration, image reconstruction, super-resolution, video compression and multiview video coding.

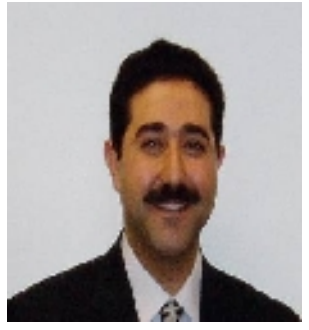

**Ahmed A. AbdelHafez**; received the B.S. and M.Sc. in Electrical Engineering from Military Technical College (MTC) in 1990, 1997 respectively, and his Ph.D from School of Information Technology and Engineering (SITE), University of Ottawa, Ottawa, Canada in 2003. Dr. Abdel-Hafez is the head of the Cryptography Research Center (CRC), Egypt where he is leading many applied researches in communication security field. He is a visiting lecturer in Communication Dept. MTC, and other universities in Egypt. Dr. Abdel-hafez published more than 40 papers in specialized conferences and periodicals. His research interests include wireless networks and data security, mathematical cryptography and provable security.